\documentclass[11pt]{article}
\usepackage{amsmath,amssymb,amsthm,amsfonts,hhline,color}

\def\H{{\bf H}}
\def\C{{\bf C}}
\def\R{{\bf R}}
\def\F{{\bf F}}
\def\Z{{\bf Z}}

\def\Q{{\bf Q}}
\def\tr{{\rm Tr}}
\def\K32{K3^{[2]}}

\def\K3n{K3^{[n]}}

\def\M24{M_{24}}
\def\M23{M_{23}}

\def\dim{{\rm dim}\,}

\def\pf{\noindent{\bf Proof:\ }}
\def\qed{\hfill\framebox[2.5mm][t1]{\phantom{x}}}

\title{The Conformal Packing Problem}

\author{Gerald H\"ohn\\
Department of Mathematics, Kansas State University\thanks{Supported by the Simons Foundation, Award ID: 355294}}

\date{September, 2019}

\begin{document}

\bibliographystyle{amsalpha}

\theoremstyle{plain}
\newtheorem{thm}{Theorem}[section]
\newtheorem{prop}[thm]{Proposition}
\newtheorem{lem}[thm]{Lemma}
\newtheorem{cor}[thm]{Corollary}
\newtheorem{rem}[thm]{Remark}
\newtheorem{conj}[thm]{Conjecture}

\newtheorem{introthm}{Theorem}
\renewcommand\theintrothm{\Alph{introthm}}

\theoremstyle{definition}
\newtheorem{defi}[thm]{Definition}

\maketitle

\abstract{We formulate the conformal packing problem and dual packing problem in analogy to similar problems for binary codes and lattices.
We obtain explicit numerical upper bounds for the minimal dual conformal weight of a unitary strongly-rational vertex operator algebra for several central charges $c$
and we also discuss asymptotic bounds.
As a main result, we find the bounds $1$ and $2$ for the minimal dual conformal weight for the central charges $c=8$ and $c=24$, respectively.
These bounds are reached by the vertex operator algebra  associated to the affine Kac-Moody algebra $\widetilde{E}_8$ at level~$1$  of central charge $c=8$
and the moonshine module  of central charge $c=24$.
The optimal bounds for these two central charges are obtained by methods similar to the one used by Viazovska and Cohn et al.\ in solving
the sphere packing problem in dimension~$8$ and $24$, respectively.}


\section{Introduction}

The sphere packing problem asks for the densest packing of balls of equal radius without overlap 
in any given dimension. This formulation applies to the Euclidean space $\R^n$ and the finite
affine space $\F_2^n$ equipped with the Hamming metric. In the case of ${\bf R}^n$ it is possible to
scale any given packing by a positive factor and so any bounds on the density depend on $n$ only.
The case of $\F_2^n$ is the one studied in coding theory: If $d$ is the minimal distance between the centers of the $A(n,d)$ balls
of the packing, one asks for bounds on the information rate $R=\log_2 A(n,d)/n$ as a function of $d/n$.

\medskip

In many ways, two dimensional conformal field theories are an analog of packings in $\F_2^n$ and  $\R^n$.
For example, they can be thought as a precise description of the physics at the event horizon of black holes
in a three-dimensional universe~\cite{3d-QG}. 
The Bekenstein bound is an upper bound for the Bekenstein-Hawking entropy of the black hole.
In recent years, upper bounds for several quantities of a conformal field theory and the corresponding black hole entropy have been found by a
technique called conformal bootstrap.

\smallskip

Mathematically, the analogy becomes even closer if one considers vertex operator algebras, which describe
the chiral parts of  conformal field theories. For example, the moonshine module of central charge $24$ is a
vertex operator algebra analog  of the Leech lattice in dimension~$24$ and the Golay code of length $24$. The moonshine module has
many remarkable properties. Its symmetries are the Monster group, the largest of the twenty-six
sporadic finite groups. It is also the starting point of the construction of the Monster Lie algebra, a construction
motivated by bosonic string theory.

The Golay code is the unique optimal solution of the packing problem in $\F_2^{24}$ for the minimal distance $d=8$ 
and the Leech lattice  is the unique optimal solution of the packing problem in  $\R^{24}$ as has been shown
recently. Thus it is natural to ask if there is a formulation of the packing problem for vertex operator algebras
of central charge~$c$ such that the Moonshine module is the unique optimal solution for central charge $24$.

\medskip

One obstacle is that there is no space known into which vertex operator algebras of central charge $c$ could be embedded.
Therefore the properties like the radius of the balls and the density of the packing have to be described intrinsically
by the properties of the vertex operator algebra. This can be done in a straightforward way.
The analog of the minimal distance between balls is the {\it minimal conformal weight\/} $\mu_V$ of a vertex operator algebra $V$.
This is the smallest positive conformal weight of a Virasoro highest weight vector.
For the analog of the density of the packing one may use the {\it inverse of the global dimension\/} of the modular tensor category associated 
to the (strongly-rational) vertex operator algebra. This can also be expressed in terms of the limit $\lim_{\tau\longrightarrow i\infty} \chi_V(\tau)/\chi_V(-1/\tau)$
where $\chi_V$ is the character of $V$. The {\it conformal packing problem\/} asks for bounds relating the minimal conformal weight and
the global dimension for given central charge.

\medskip

Another issue is that vertex operator algebras are the analog of even lattices or doubly-even linear codes. Only full conformal
field theories more closely  resemble lattices or linear codes. In particular, the smallest possible global dimension
is~$1$ in which case the vertex operator algebra is self-dual. For self-dual  vertex operator algebras,
I showed the upper bound $[c/24]+1$ for the minimal conformal weight~\cite{Ho-dr}. Together with the self-duality of the Moonshine module this already
shows that the moonshine module is an optimal solution for the conformal packing problem if one fixes either its minimal conformal weight
or its global dimension.  Although the general conformal packing problem remains interesting, nothing new can be said
by it for the moonshine module.

\medskip

The following approach overcomes this problem. For a lattice $L$ or linear code $C$, one can define the dual lattice $L^*\subset \R^n$ or dual code $C^\perp\subset \F_2^n$.
Even if the lattice $L$ is even or the code $C$ is doubly-even, the dual lattice $L^*$ and dual code $C$ have not to be even or doubly-even, respectively.
For vertex operator algebras $V$, one may define a dual $V^*$ as the direct sum of representatives of the equivalent classes of irreducible $V$-modules.
One can now consider the {\it minimal dual conformal weight\/} $\mu_V^*$ of $V^*$ which is  the smallest positive conformal weight of a Virasoro highest weight vector in $V^*$.
The {\it conformal dual packing problem\/} asks for bounds for $\mu_V^*$  for given central charge~$c$. If we are interested in upper bounds, we may ignore 
the global dimension since it is automatically bounded from below by~$1$.

\medskip

The analogous {\em dual packing problems\/} for even lattices and doubly-even codes are non-trivial. We will show that the Leech lattice is the 
optimal solution for the dual lattice packing problem in dimension $24$ and the Golay code is the solution for the dual linear code packing problem for codes
of length $24$. The proof for dimension~$24$ is almost as hard as the general packing problem in that dimension and requires
the use of ``magic functions'' of the type first constructed by Viazovska.

\medskip

For vertex operator algebras, we will show that the moonshine module is an optimal solution of the dual conformal packing problem for central charge $c=24$.
We also find a sharp bound for vertex operator algebras of central charge $8$ where the vertex operator algebra associated to the affine Kac-Moody algebra $\widetilde{E}_8$ at level~$1$
provides the optimal solution of the conformal dual packing problem.

For other fixed values of the central charge $c$ we  compute numerically an upper bound for $\mu_V^*$ 
by using techniques from semidefinite programming.

\medskip

Our approach uses the conformal bootstrap method, and combines it with the magic function approach. It is motivated
by earlier numerical computations of the author regarding properties of the modular $j$-function under the modular transformation $\tau\mapsto -1/\tau$ and the papers~\cite{He, FK, CLY, HMR}.
The approach of the last paper is very similar, but by using full conformal field theories instead of vertex operator algebras the
bounds obtained there are not sharp. Another motivation for studying the conformal packing problem is to investigate properties ``dual'' to the properties of
{\it conformal designs\/} as introduced in~\cite{Ho-designs}.

\bigskip

The paper is organized as follows. In the next two sections we will discuss the dual packing problem for binary codes and lattices.
The point of those section is mainly to compare the obtained results for vertex operator algebras with the analogous results for
codes and lattices.
In Section 4, we will give a more detailed formulation of the conformal packing problem and its dual version.
In Section 5, we numerically compute upper bounds for the minimal dual conformal weight for fixed central charges.
In the last section we will find the optimal upper bound for central charges $8$ and $24$.

\bigskip

{\bf Acknowledgments.} I am grateful to Christoph Keller for bringing the paper~\cite{HMR} to my attention. 
I would like to thank Richard Borcherds, Chongying Dong, Terry Gannon, Christoph Keller, Xiannan Li, Geoffrey Mason, Sven M\"oller and  Gabriele Nebe for discussions.
I would also like to thank Philipp H\"ohn and Sven M\"oller for proofreading the manuscript.


\section{The dual packing problem for binary codes}

In this section we discuss bounds for the minimum dual Hamming weight of doubly-even codes.
We also illustrate the analog of our method which we will use for obtaining upper bounds for the lattice case and the conformal dual packing problem.

\bigskip

A {\it binary code\/} $C$ is a subset of a finite dimensional vector space over the field $\F_2$ with two elements.
The dimension of the vector space is called the length of $C$. In the following, all codes are assumed to be binary codes.

The orthogonal or {\it dual code\/} of a code $C$ of length $n$ is defined by
$$C^\perp=\{d\in \F_2^n \mid (c,d)=\sum_{i=1}^n d_i\cdot c_i=0 \ \hbox{for all}\ c \in C\}.  $$

The  {\it weight\/} ${\rm wt}(c)$ of a vector $c=(c_1,c_2,\ldots)$ is the number of non-zero coordinates~$c_i$. 
A code $C$ is called {\it doubly-even\/} if it is a linear subspace and the  weights
of all vectors $c$ in $C$ are divisible by $4$.
For a doubly-even code $C\subset\F_2^n$ one has $C\subset C^\perp$.
Doubly-even codes are the coding theoretic analog of even lattices and vertex operator algebras.

We denote by $\mu_C=\min(\{{\rm wt}(c)\mid c\in C\setminus\{0\}\})$ the {\it minimum weight\/} of $C$ and by
$\mu^*_C=\mu_{C^\perp}$ the {\it minimum dual weight.\/} For doubly-even codes one has $\mu_C^*\leq \mu_C$.

\smallskip

Codes $C$ satisfying $C=C^\perp$ are called {\it self-dual.}\ \
The length $n$ of a self-dual doubly-even code has to be divisible by $8$
and for its minimal weight one has the upper bound  $\mu_C\leq 4[n/24]+4$ 
where $[\ ]$ denotes the Gaussian bracket~\cite{MaSl}.

\medskip

We explain first how to find an upper bound for $\mu_C^*$ using a method similar to the one we will use for
vertex operator algebras. Although this bound will in general be weaker than the best known, it agrees with 
it for several smaller values of $n$ including $8$ and~$24$.

\smallskip

The weight enumerator of a linear binary code $C$ of length $n$ is the bivariate polynomial
$$W_C(x,y)=\sum_{c\in C}x^{{\rm wt}(c)}y^{n-{\rm wt}(c)}.$$
The {\it MacWilliams identity\/} computes the weight enumerator of the dual code in terms of
the weight enumerator of the code:
$$ W_{C^\perp}(x,y)={\textstyle \sqrt{|C^\perp/C|}}\   W_C\left(\frac{y-x}{\sqrt{2}},\frac{x+y}{\sqrt{2}}\right).$$

The matrix $S=\frac{1}{\sqrt{2}}\left(\begin{smallmatrix}-1 & 1 \\ \phantom{-}1 & 1 \end{smallmatrix}\right)$
acts on the polynomial ring $\R[x,y]$ via the substitution $\left(\begin{smallmatrix}x\\  y\end{smallmatrix}\right) \mapsto
S\left(\begin{smallmatrix}x \\y\end{smallmatrix}\right)$. We write this action by $p|S$ for a polynomial $p$ in $\R[x,y]$.
The square $S^2$ is the identity matrix. One easily finds that the ring
$\R[x,y]^S$ of polynomials invariant under the action of $S$ is algebraically freely generated
by the two homogeneous polynomials $A=(\sqrt{2}-1)\,x+y $ and $B=x^2+y^2$ of degree $1$ and $2$, respectively.

The following result follows directly from the MacWilliams identity.
\begin{lem}
Let $C$ be a linear code of length $n$ and let
$$F_C=\frac{1}{2}\left(W_C+W_C|S\right)=\sum_{i=0}^n c_i\,x^iy^{n-i}.$$
Then $F_C$ is a polynomial in $\R[x,y]^S= \R[A,B]$ with non-negative coefficients $c_i$ and one has $c_0>0$.
If~$C$ has minimum weight $\mu_C$ and minimum dual weight $\mu^*_C$ then $c_i=0$ for $i=1$, $\ldots$, \hbox{$\min(\mu_C,\mu^*_C)-1$.}
\end{lem}

The $S$-invariant polynomial $F_C$ can be written in the form $F_C=\sum_{j=0}^{[n/2]} b_j A^{n-2j}B^j$ with real numbers $b_j$ for $j=0$, $\ldots$, $[n/2]$.
Its coefficients $c_i$ are then linear combinations of the $b_j$.  We consider the linear programming problem in
the variables $b_0$, $\ldots$, $b_{[n/2]}$ maximizing the (arbitrarily chosen) objective function $\sum_{i=0}^nc_i$  subject to  the linear constraints $c_i=0$ for $i=1$, $\ldots$, $k$,
$c_i\geq 0$ for $i=k+1$, $\ldots$, $n$ and the chosen normalization $c_0=1$. If this problem is infeasible, i.e.\ the constraints are inconsistent,
then $k$ must be an upper bound for $\min(\mu_C,\mu^*_C)$. If $C$ is further assumed to be doubly-even then $k$ is an upper bound for $\mu^*_C$ alone.

We computed the optimal upper bound one can get with this method for various lengths $n$.
\begin{thm}
For the values of $n$ listed in Table~\ref{codebounds}, the row $\mu$ provides an upper bound
for $\min(\mu_C,\mu_C^*)$ for any linear code $C$ of length $n$. 
\end{thm}
\begin{table}\caption{Linear programming bounds for $\min(\mu_C,\mu_C^*)$}\label{codebounds}
$$\begin{array}{c|*{12}{c}}
n   & 0 & 1 & 2 & 3 & 4 & 5 & 6 & 7 & 8 & 9 & 10 & 11 \\ \hline
\mu & - & 1 & 2 & 2 & 2 & 2 & 3 & 3 & 4 & 4 & 4  & 4  \\
\\
n   & 12 & 13 & 14 & 15 & 16 & 17 & 18 & 19 & 20 & 21 & 22 & 23 \\ \hline
\mu &  4  & 4 & 5  & 5  &  6 &  6 &  6 &  6 &  6 & 6  &  7 &  7 \\
\\
n   & 24 & 25 & 26 & 27 & 28 & 29 & 30 & 31 & 32 & 33 & 34 & 35 \\ \hline
\mu &  8 &  8 & 8  & 8  &  8 &  8 & 8  &  8 & 9  & 9  & 10 & 10 \\
\\
n   & 40 & 48 & 56 & 64 & 72 & 80 & 88 & 96 & 104 & 112 & 120 & 128 \\ \hline
\mu & 10 & 12 & 14 & 16 & 18 & 19 & 21 & 23 &  24  \\
\end{array}
$$ 
\end{table}

We also note that for even $n$ the maximal achieved value for $\sum_{i=0}^nc_i$ is $2^{n/2}$ suggesting that a code $C$ meeting the bound
must be self-dual. However, the corresponding polynomial $F_C$ which would be equal to $W_C$ in this case has usually non-integral coefficients
indicating that such a code probably does not exist in those cases.

\medskip

For double-even codes, the value for $\mu$ in Table~\ref{codebounds} is an upper bound for $\mu_C^*$.
For those codes, a bound almost as strong as the bound $4[n/24]+4$ for doubly-even self-dual codes is known.
\begin{thm}[Rains~\cite{rains}]
Let $n=24k+l$ with integral $k$ and $0\leq l < 24$. Then the minimum dual weight of a doubly-even code of length~$n$ is bounded
from above by $4[n/24]+d_l$ where the numbers $d_l$ are given by the following table:
$$\begin{array}{c|*{12}{c}}
l   & 0 & 1 & 2 & 3 & 4 & 5 & 6 & 7 & 8 & 9 & 10 & 11 \\ \hline
d_l & 4 & 1 & 1 & 1 & 2 & 1 & 2 & 3 & 4 & 1 &  2 &  2 \\
\\
l   & 12 & 13 & 14 & 15 & 16 & 17 & 18 & 19 & 20 & 21 & 22 & 23 \\ \hline
d_l & 2 &  2 & 3 & 3 & 4 & 4 & 4 & 3 & 4 & 5 & 6 &  7
\end{array}
$$ 

\end{thm}
One may expect a bound of such a type since one can assume that a doubly even code  $C$ for which the maximal possible $\mu_C^*$ is achieved is maximal
in the sense that it is not the subspace of a larger doubly-even code. This forces $C$ to be ``almost self-dual'' in the sense that
$C^\perp/C$ is small.

Our linear programming bound agrees for doubly-even codes with the Rains bound for $n=1$, $2$, $3$, $4$, $7$, $8$, $23$ and $24$ and otherwise it seems weaker.
Our bound for all the considered values of $n$ is at least as strong as the bound $n/4+2$ and we expect that this will be true in general.
Since the Rains bound is at least as strong, we will not investigate this here further. 
It is likely  that by studying the coefficients of $W_C$ and $W_{C^\perp}$ individually, stronger linear programming bounds
can be obtained.

\medskip

We can compare the asymptotic Rains bound $n/6+O(1)$ and our conjectured bound $n/(4+\epsilon)$ with the best known general
lower and upper asymptotic bounds from coding theory~\cite{McSl}. If we assume that the information rate $R$ is $1/2$ which corresponds to
$|C^\perp|\approx |C|\approx 2^{n/2}$,
the McEliece-Rodemich-Rumsey-Welch asymptotic bound gives $n/5.3454...$ as an upper asymptotic bound for $\mu^*_C$ whereas the rather simple 
Gilbert-Varshamov bound gives $n/9.0885...$ as an asymptotic lower bound for $\sup_C \mu^*_C$. 
It is known that there exists doubly-even self-dual codes which meet the Gilbert-Varshamov bound asymptotically.

\medskip

Finally we mention that it is well known that our bounds $\mu_C^*\leq 4$ for codes of length $8$ and $\mu_C^*\leq 8$ for codes of length $24$
are met by the {\it extended  Hamming code ${\cal H}_8$  of length $8$\/} and the {\it extended Golay code ${\cal G}_{24}$ of length $24$,\/} respectively.
The linear programming method determines the weight enumerators of any codes meeting those bounds uniquely.
From this it follows that the codes are must be self-dual and this allows to show that ${\cal H}_8$ and ${\cal G}_{24}$
are the unique doubly-even codes with that property.


\section{The dual packing problem for lattices}

In this section we discuss bounds for the minimal dual norm of even lattices. 
We show how to  use semidefinite programming to obtain explicit upper bounds for selected dimensions.
For dimensions $8$ and $24$, sharp upper bounds follow from the solution
of the sphere packing problem in those dimensions.

\bigskip

A {\it lattice \/} $L$ is a discrete subgroup of $\R^n$ of maximal rank. 
As abelian group one has $L\cong\Z^n$.

The {\it dual lattice\/} of a lattice $L\subset \R^n$ is the lattice
$$L^*=\{y\in \R^n \mid (x,y)\in \Z \ \hbox{for all}\ x \in L\}.  $$

The {\it norm\/} $|x|^2$ of a vector $x\in \R^n$ is its squared Euclidean length.
A lattice is called {\it even\/} if the norm of all vectors $x$ in $L$ are even.
For an even lattice $L$ one has $L\subset L^*$.

We denote by $\mu_L=\min(\{|x|^2\mid x\in L\setminus\{0\}\,\})$ the {\it minimal norm\/} of $L$ and by
$\mu^*_L=\mu_{L^*}$ the {\it minimal dual norm.\/} For even lattices one has $\mu_L^*\leq \mu_L$.

\smallskip

Lattices $L$ satisfying $L=L^*$ are called  self-dual or {\it unimodular.}\ \
The rank of an even unimodular lattice has to be divisible by $8$
and for its minimal norm one has the upper bound  $\mu_L\leq 2[n/24]+2$~\cite{CoSl}. 

\medskip

The theta series of a lattice $L\subset \R^n$ is the series
$$\theta_L=\sum_{x\in L } q^{\frac{1}{2}|x|^2}.$$
If we set $q=e^{2\pi i \tau}$ where $\tau$ is in the complex upper half-plane  $\H$, then the series converges locally uniformly and absolutely to
a holomorphic function.
The Poisson summation formula can be used to relate the theta series of the dual lattice with the
the theta series of the lattice:
$$ \theta_L(-\frac{1}{\tau})=\frac{1}{{\rm vol}(\R^n/L)}\, \left(\frac{\tau}{i}\right)^{n/2}\,\theta_{L^*}(\tau) .$$
Here, ${\rm vol}(\R^n/L)$ denotes the volume of a fundamental parallelotope of $L$.

The matrix $S=\left(\begin{smallmatrix}0 & -1 \\ 1 & \phantom{-}0 \end{smallmatrix}\right)$
acts on the upper half-plane $\H$  via the M\"obius transformation $\tau \mapsto S\tau=-\frac{1}{\tau}$. 
We use the notation $f|S$ for the induced action on functions $f$ on $\H$.

\medskip

The following result follows directly from the above transformation property of the theta series of a lattice
under the action of $S$. 
\begin{lem}
Let $L\subset \R^n$ be a lattice and let
$$F_L=\frac{1}{2}\left(\frac{\tau}{i}\right)^{n/4}\left(\theta_L+\theta_L|S\right)=\sum_{h\geq 0} c_h\,\left(\frac{\tau}{i}\right)^{n/4}q^h.$$
Then one has $F_L=F_L|S$ and $F_L$ has the given expansion with non-negative coefficients $c_h$ and $c_0>0$.
If~$L$ has minimal norm $\mu_L$ and minimal dual norm $\mu^*_L$ then $c_h=0$ for all $h$ in the 
open interval from $0$ to \hbox{$\min(\mu_L,\mu^*_L)$.}
\end{lem}

In the following, we will show that these properties of $F_L$ are enough to obtain upper bounds for $\min(\mu_L,\mu^*_L)$.

\medskip

We restrict $\tau$ to the positive imaginary axis and write $\tau=i\,t$ with positive real $t$.
Then $\overline{F}_L(t)=F_L(it)$ is a real-valued and real analytic function and satisfies 
$$\overline{F}_L\left(\frac{1}{t}\right)=\overline{F}_L(t).$$

Since $(t\,\partial_t)^m\left(\frac{1}{t}\right)=(-1)^m\,\frac{1}{t}$ we see that the ``odd'' functionals
$${\cal D}[f]=\sum_{m=0}^N a_{2m+1}\,(t\,\partial_t)^{2m+1}(f)\bigm\vert_{t=1}$$ 
annihilate $\overline{F}_L$, i.e.\, ${\cal D}(\overline{F}_L)=0$.

We study the space of such functionals.
For the individual terms in the expansion of $\overline{F}_L$ one has
$$(t\,\partial_t)^m\left(t^{n/4}e^{-2\pi ht}\right)\Bigm\vert_{t=1}=p_m(h)\,e^{-2\pi h}$$
with a real polynomial $p_m$ of degree~$m$
and so ${\cal D}\left[t^{n/4}e^{-2\pi ht}\right]=p(h)\,e^{-2\pi h}$ where $p=\sum_{m=0}^N a_{2m+1}\,p_{2m+1}$ is a polynomial of degree $2N+1$.

Assume we can find real numbers $a_{2m+1}$, $m=0$, $\ldots$, $N$, such that $p(0)>0$ and $p(h)\geq 0$ for $h\geq \Delta$.
Then one has
$$0 ={\cal D}[\overline{F}_L]=c_0\,p(0)+\sum_{h>0 } c_h\,p(h)\,e^{-2\pi h}.$$
Since $c_0\,p(0)$ is positive and $c_h\,p(h)\,e^{-2\pi h}$ is non-negative for $h\geq \Delta$, at least one of the $c_h$ for  $0<h< \Delta$
has to be nonzero. Thus we have found an upper estimate $\min(\mu_L,\mu^*_L)<\Delta$ which must hold for all lattices $L$.

\smallskip

This polynomial optimization problem can be rewritten as a semidefinite programming problem for which powerful numerical solvers exist.
We used the solver SDPB~\cite{sdpb} since it allows us directly to use the polynomial formulation of the problem as input.
We have chosen $p(0)$ as the objective function to be maximized and $p(x+\Delta)\geq 0$ for $x\geq 0$ as the polynomial constrained.
For fixed dimension $n$, the approximately smallest values of $\Delta$ for which a positive solution for the objective function exists seem to converge 
for increasing degree $2N+1$ of the polynomial $p$. The results of the computation for selected dimensions are shown in 
Table~\ref{latticebounds}.~\footnote{For ranks $1$ to $4$, the chosen particular form of the semidefinite programming problem seems
to have either no unique solution or it is numerically instable.}

\begin{table}\caption{Semidefinite programming bounds for $\min(\mu_L, \mu_L^*)$}\label{latticebounds}
$$\begin{array}{c|*{8}{c}}
n    & 5     & 7     & 8                 & 16    &  23    & 24          & 32    & 48    \\ \hline
N=12 & 1.591 & 1.866 & 2.00018           & 3.028 &  3.892 & 4.014       & 4.986 & 6.932 \\
N=24 & 1.591 & 1.866 & 2.0000005         & 3.026 &  3.881 & 4.00024     & 4.949 & 6.802 \\
N=36 & 1.591 & 1.866 & 2+5 \cdot 10^{-9} & 3.026 &  3.880 & 4.000007    & 4.947 & 6.792
\end{array}
$$ 
\end{table}

We summarize this as follows:
\begin{thm}
For lattices $L$ in dimensions $n=5$, $7$, $8$, $16$, $23$, $24$, $32$ and $48$ the minimum of the minimal norms of $L$ and $L^*$ 
has the upper bound given by the entry in row $N=36$ of Table~\ref{latticebounds}.
\end{thm}

\medskip

We can compare our method for an upper bound for $\min(\mu_L,\mu^*_L)$ with general sphere packing bounds.

\smallskip

One  can always rescale a lattice $L$  by a positive factor such that one has $\mu_L=\mu^*_L$ for the rescaled lattice. 
For this scale factor $\min(\mu_L,\mu^*_L)$ has the largest possible value.
Either ${\rm vol}(\R^n/L)$ or ${\rm vol}(\R^n/L^*)$ (or both) are less than or equal $1$ and we may apply any general 
sphere packing bound to either $L$ or $L^*$ to obtain an upper bound for $\mu_L=\mu^*_L=\min(\mu_L,\mu^*_L)$.

In~\cite{HMR}, the authors used an approach similar as above and found the asymptotic upper bound $n/8.85629...$. 
The asymptotic Kabatiansky-Levenshtein bound is the best known asymptotic upper sphere packing bound and provides the 
better upper asymptotic bound $n/9.795...$ for $\min(\mu_L,\mu^*_L)$.

\smallskip

One also can always scale a lattice $L$ by a positive factor such that ${\rm vol}(\R^n/L)={\rm vol}(\R^n/L^*)=1$. 
But only if we assume that  $L$ and $L^*$ are then isometric or at least satisfy $\mu_L=\mu^*_L$, 
a bound for $\min(\mu_L,\mu^*_L)$ will also provide a bound for $\mu_L$. 
However, in practice it seems always 
possible to get a similar good linear programming bound for general sphere packings and thus the problem to find upper bounds for the the dual
minimal norm for arbitrary lattices is only slightly different from the general lattice packing problem.  

\smallskip

In summary, the above semidefinite  programming approach does not seem to achieve better upper bounds 
than the general sphere packing upper bounds and we therefore considered only a few sample dimensions.

\bigskip

We are most interested in even lattices since this is the analogous case when compared with vertex operator algebras.
Then our bounds for $\min(\mu_L,\mu^*_L)$ are bounds for the minimal dual norm $\mu^*_L$.

\smallskip

For finding upper bounds for $\mu_L^*$ for even lattices, we can restrict ourselves to lattices which are maximal
in the sense that they are not a sublattice of a larger even lattice of the same rank. An even lattice $L$
is maximal if and only if there is no non-zero isotropic vectors in its discriminant group $L^*/L$ with respect
to the induced quadratic form.

\medskip

For even lattices, the upper bound can often be improved.

Recall that the {\it level\/} of an even lattice $L$ is the exponent of the discriminant group $L^*/L$.
In the paper~\cite{BoeNe}, the authors consider maximal even lattices of even rank $n=2k$ and level $N$. They
show that the theta series of the rescaled dual lattice $\sqrt{N}L^*$ for square free levels $N$ with $L^*/L=N^2$ belongs to a certain 
subspace $M_k(N)^*$ of the space of modular forms of weight $k$ for the congruence subgroup $\Gamma_0(N)$.
Furthermore, there is a unique {\it extremal modular form\/} $F$ in $M_k(N)^*$, which means that the constant coefficient
of $F$ is $1$ and the next $\dim M_k(N)^*-1$ coefficients are zero.
They also give for even $k\geq 2$ the following dimension formula for $M_k(N)^*$:
$$\dim  M_k(N)^*=\frac{(k-1)N}{12}+\frac{1}{2}-\frac{1}{4}\left(\frac{-1}{(k-1)N}\right)-\frac{1}{3}\left(\frac{-3}{(k-1)N}\right).$$
Thus for large $k$ one has $\dim  M_k(N)=\frac{Nk}{12}+O(1)$. If one considers the unscaled lattice $L^*$ 
the first nonzero coefficients besides the first should occur around the norm $n/12+O(1)$.
However, this does not translate directly into a bound for $\mu_L^*$ for those lattices since it can happen and in some cases it does indeed
happen that further coefficients of the extremal modular form vanish. Nevertheless, this result provides a strong
indication that for even lattices the asymptotic upper bound  $n/12+O(1)$ for $\mu_L^*$ may hold in general.

\smallskip

On the other hand, it is known that even unimodular lattices exist which meet the asymptotic Minkowski
bound $n/(2\pi e)=1/17.07946\ldots $,~cf.~\cite{CoSl}.

\bigskip

It follows from the solution of the sphere packing problem in dimension $8$ and $24$ or the slightly simplified estimates
for even lattices that one has the upper bounds $\mu_L^*\leq 2$ for even lattices of rank~$8$ and $\mu_L^*\leq 4$ for even lattices of rank~$24$.
Those bounds are met by the {\it root lattice $E_8$ of rank~$8$} and the {\it Leech lattice of rank $24$,\/} respectively.
One also sees from the proof that the dual $L^*$ of any lattice $L$ meeting those two bounds must be even, i.e.\ $L$ has to be even unimodular.
It is well-known, cf.~\cite{CoSl}, that both lattices are the unique lattices with that property.


\section{Conformal packing and dual packing problems}

We are interested in vertex operator algebras which have a modular tensor category as representation category
and which satisfy a unitary assumption.

\medskip

For the first condition it is by results of Huang~\cite{Huang-tensor} sufficient to assume that the vertex operator algebras $V$ satisfy the following conditions:
\begin{itemize}
\item $V$ is of CFT-type, i.e.\ one has $V=\bigoplus_{n\in\Z_{\geq 0}} V_n$ with respect to the $L_0$-grading and $V_0=\C\,{\bf 1}$;
\item $V$ is  rational, i.e.~the module category is semisimple;
\item $V$ is  $C_2$-cofinite;
\item $V$ is simple;
\item $V$ is self-contragredient, i.e.\ $V$ is isomorphic to its contragredient module $V'$.
\end{itemize}
We assume that the vertex operator algebras $V$ in this paper satisfy these assumptions which we call {\it strongly-rational\/}.
From the rationality assumptions it follows that $V$ has only finitely many
equivalence classes of irreducible modules.

\smallskip

We also like to make a unitarity assumption on $V$. Since the existence of an invariant hermitian form on $V$ and its modules
seems in general  to require additional data, we replace it by an assumption involving only the Virasoro algebra.

A vertex operator algebra is a module of the Virasoro algebra of a certain central charge $c$.
For $c$, $h\in \C$ we denote by 
$M(c, h)$ the highest weight module for the Virasoro algebra with central charge $c$ and highest weight $h$. 
We set $\overline{M}(c, 0) = M(c, 0)/U({\rm Vir})L_{-1} v$
where $v$ is a highest weight vector with highest weight $0$.
A Virasoro algebra module is called unitary if there exists a positive definite hermitian form such that the adjoint of 
$L_{n}$ is $L_{-n}$. If $c$ and $h$ are real and $c>1$, the modules $M(c, h)$ for $h>0$ and  $\overline{M}(c, 0)$ are irreducible and unitary.
For central charges $c>1$ it is known that a strongly-rational vertex operator algebra $V$ is completely reducible as a module for the Virasoro algebra~\cite{DJ}.

We are interested in vertex operator algebras $V$ for which similar properties hold for all $V$-modules:
\begin{itemize}
\item Every irreducible $V$-module is completely reducible as a module for the Virasoro algebra.
\item The conformal weights of the irreducible $V$-modules non-equivalent to $V$ are positive.
\end{itemize}
We call such a vertex operator algebra {\em unitary.} The last condition also guarantees that
$\overline{M}(c, 0)$ occurs only in the irreducible $V$-modules isomorphic to $V$ and has for those
modules the multiplicity~$1$. 

\bigskip

We recall the following definition from~\cite{Ho-dr}.
\begin{defi}
The {\em minimal conformal weight\/} $\mu_V$ of a vertex operator $V$ is the smallest conformal weight of any Virasoro highest weight vector in $V$
different from the vacuum vector ${\bf 1}$.
\end{defi}

\smallskip
In analogy to the packing problems in $\F_2^n$ and ${\bf R}^n$, one may hope that unitary strongly-rational vertex operator algebras $V$ of an arbitrarily large  
minimal conformal weight $\mu_V$ may  exist for large enough central charges $c$.
More generally, one may expect that for fixed $c$ there exist upper bounds for $\mu_V$  depending on the global dimension $D(V)$ of the modular tensor category associated to the vertex operator algebra. 
The {\em general conformal packing problem\/} asks therefore for the possible values of
$$(c,\mu_V,D(V)) \in {\bf Q}_{>0} \times {\bf Z}_{\ge 1} \times \overline{{\bf Q}}_{\geq 1}$$
which can be realized by unitary strongly-rational vertex operator algebras.

Since the global dimension $D(V)$ is the analog of the square of the volume of the fundamental domain of a lattice packing, one may ask
more specifically for the relation between the quantities
$$  \frac{\mu_V}{c} \qquad \hbox{and} \qquad -\frac{\log(D(V))}{c}   $$
for fixed $c$ and for the limit $c\longrightarrow \infty$ and for bounds similar to the 
Gilbert-Varshamov and McEliece-Rodemich-Rumsey-Welch bounds for codes~\cite{McSl} and the Minkowski and 
Kabatiansky-Levenshtein bounds for sphere packings~\cite{CoSl}.

\smallskip
The  global dimension $D(V)$ has to be at least $1$. This corresponds to the fact that 
vertex operator algebras are the analog of doubly-even and even lattices and that there is no notion of non-integral vertex operator algebras.

\medskip

\begin{defi}
For a vertex operator algebra $V$ a {\em dual of $V$}, denoted by $V^*$, is any $V$-module
$V^*=\bigoplus_{\lambda \in \Lambda}  M_\lambda$
where $\Lambda$ is a set which indexes the inequivalent irreducible $V$-modules and
$M_\lambda$ is a corresponding $V$-module.
\end{defi}

A vertex operator is called {\em self-dual\/} or holomorphic if $V\cong V^*$ as $V$-module.
Self-dual strongly-rational vertex operator algebras can only exist for central charges $c$ divisible by $8$.
The following theorem was proven in~\cite{Ho-dr}.
\begin{thm}
For a self-dual unitary strongly-rational vertex operator algebra $V$ one has the bound
$$\mu_V\leq \left[\frac{c}{24}\right]+1.$$
\end{thm}
Self-dual vertex operator algebras $V$ for which $\mu_V$  meets this upper bound are called {\em extremal.}
Extremal self-dual vertex operator algebras are so far only known for central charges $c=8$, $16$, $24$, $32$ and $40$, see~\cite{Ho-dr}.
The affine Kac-Moody algebra $\widetilde{E}_8$ at level~$1$~\cite{FZ} and the moonshine module~\cite{Bo-ur,FLM}
are examples of extremal vertex operator algebras of central charge $c=8$ and $c=24$, respectively.

\smallskip

A unitary strongly-rational vertex operator is self-dual if and only if its global dimension $D(V)$ is equal to $1$.
It follows that for  $D(V)=1$ and central charges $c=8$ and $c=24$ the minimal weight is bounded from above by $1$ and $2$, respectively,
and these two bounds are sharp. To obtain for these central charges an interesting and new problem,
we consider the minimal dual conformal weight instead.

\begin{defi}
The {\em minimal dual conformal weight\/} $\mu_V^*$ of a vertex operator $V$ is the smallest conformal weight of any Virasoro highest weight vector in $V^*$
different from the vacuum vector ${\bf 1}$.
\end{defi}

\smallskip
Since for self-dual vertex operator algebras one has $\mu^*_V=\mu_V$, the following corollary holds.
\begin{cor}
For a self-dual unitary strongly-rational vertex operator algebra $V$ one has the upper bound \hbox{$\mu^*_V\leq [c/24]+1$}.
\end{cor}

\medskip

For self-dual vertex operator superalgebras $W$, the weaker bound $\mu_W\leq [c/16]+\frac{1}{2}$ for the minimal weight
was also given in~\cite{Ho-dr}.
In~\cite{Ho-extremal}, it was  shown that the bound can be improved to $\mu_W\leq  [c/24]+1$ for all
$c\not= 23\frac{1}{2}$. For $c= 23\frac{1}{2}$ the bound is $\frac{3}{2}$ which is obtained by the 
shorter Moonshine module $V\!B^{\natural}$.

The even vertex operator subalgebra $V$ of a self-dual vertex operator superalgebra $W$ has three or four non-equivalent irreducible modules.
Since $W$ is the direct sum of $V$ and one of the other modules one has $\mu_V^*\leq \mu_W$ and hence we also obtain an upper bound for the minimal dual conformal weight.
\begin{thm}
Let $V$ be the even vertex operator subalgebra of a self-dual unitary strongly-rational vertex operator superalgebra of
central charge $c\not= 23\frac{1}{2}$. Then one has 
$$\mu^*_V\leq  \left[\frac{c}{24}\right]+1.$$
\end{thm}
For the shorter Moonshine module one has \hbox{$\mu^*_V=\frac{3}{2}$.}

\medskip

We are interested in finding the optimal  upper bound for $\mu_V^*$ which is satisfied
for all  unitary strongly-rational vertex operator algebras of fixed central 
charge $c>1$. We call this the {\em conformal dual packing problem.\/}

\smallskip

A basic observation is that it would be enough to consider only {\em maximal\/} vertex operator algebras $V$ in the sense that no 
vertex operator algebra extension of $V$  by a $V$-module exist.
Indeed,  any  module of an extension $W$ of $V$ is a  direct sum of certain $V$-modules and not all $V$-modules 
are required to occur inside a $W$-module.

\bigskip

In the next sections, we will use the semidefinite programming method 
to obtain upper bounds for $\mu_V^*$ for certain values of the central charge.
To this end, we have to investigate functions derived from the character of the vertex operator algebra.

\smallskip

The character of a $V$-module  $M$ is defined as the series
$$\chi_M=\tr(q^{L_0-\frac{c}{24}} \vert M)=\sum_{h} \dim M_h\, q^{-\frac{c}{24}+h}$$
where $c$ is the central charge of $V$.
We let $q=e^{2\pi i \tau}$ with $\tau$ in the complex upper half-plane $\H=\{z \in \C \mid \Im(z) > 0 \}$. 
\begin{thm}[Zhu~\cite{Zhu-modular}]\label{Zhu-modular}
For a unitary strongly-rational vertex operator algebra~$V$ the character $\chi_M(\tau)$ of a $V$-module $M$ converges to a holomorphic function on the upper half-plane.
The vector space spanned by the characters of the irreducible modules is
invariant under the action induced by the action of the modular group $\Gamma={\rm PSL}(2,\Z)$ on the upper half-plane. 
\end{thm}
If one takes the vector space spanned by the genus one $1$-point correlation functions instead of the characters only, one obtains a representation $\rho$
of ${\rm SL}(2,\Z)$ of dimension equal to the number of irreducible $V$-modules.
We denote by  $\Lambda$ a set which indexes the inequivalent irreducible $V$-modules with
$M_\lambda$ a $V$-module corresponding to $\lambda \in \Lambda$.
One usually takes $\Lambda=\{0,\,1,\, \ldots\, n-1\}$ with $M_0\cong V$.

Recall from the last section that the matrix $S=\left(\begin{smallmatrix}0 & -1 \\ 1 & \phantom{-}0 \end{smallmatrix}\right)$ 
acts on the upper half-plane and on functions on $\H$. 
We define the matrix $(S_{i,j})_{i,\,j\in \Lambda}$ by
$$S_{i,j}=\rho(S)_{i,j}, \qquad i, \, j \in \Lambda.$$
The quantum dimension $m_M$ of a module $M$ is defined as the limit
$$m_M=\lim_{t \longrightarrow 0} \frac{\chi_M(it)}{ \chi_V(it)}$$
where $t$ is real and positive. 

The following result is easily obtained from the modular transformation properties of the characters.
\begin{thm}
For a unitary strongly-rational vertex operator algebra 
the quantum dimensions are positive real numbers and one has
$$m_{M_i}=\frac{S_{i,0}}{S_{0,0}}$$ 
for $i\in \Lambda$.
\end{thm}
It follows from the above mentioned result of Huang that the matrix $(S_{i,j})_{i,\,j\in \Lambda}$ is equal to the
$S$-matrix defined for the modular tensor category of $V$.
Furthermore, one has 
$$D(V)=\sum_{i \in \Lambda} m_{M_i}^2=\frac{1}{S_{0,0}^2}$$ for the {\it global dimension of $V$.} 

The main property of the character  which we need is the behavior of $\chi_V$ under the action of $S$.
\begin{prop}
The character of a unitary strongly-rational vertex operator algebra $V$ satisfies
$$\chi_V(\tau \vert S)=\chi_V(-\frac{1}{\tau})=\sum_{\lambda\in\Lambda} a_\lambda \, \chi_{M_\lambda}(\tau)$$
with non-negative real numbers  $a_\lambda$.
\end{prop}
\pf The entries $S_{0,i}=m_{h_i}\, \cdot S_{0,0}$ of the matrix $\rho(S)$ are positive real numbers. \qed 

\smallskip

Since $S^2$ is the identity in $\Gamma$ we have the following result.
\begin{thm}\label{positive}
For a unitary strongly-rational vertex operator algebra $V$ of central charge~$c$, the function 
$$F_V=\frac{1}{2}\left(\chi_V+\chi_V|S\right)=c_0\,q^{-\frac{c}{24}}+\sum_{h>0} c_h\,q^{-\frac{c}{24}+h}$$ is an 
$S$-invariant function on the upper half-plane
with a $q$-expansion having coefficients $c_h\geq 0$ and $c_0>0$. 
Furthermore, the number of non-zero $c_h$ for $h$ in an interval $[0,u]$ is bounded by a linear function in $u$.
\end{thm}
This positivity property will be enough to obtain sharp bounds for $c=8$ and $c=24$.

We will also need an upper estimate for the coefficients $c_h$.
\begin{thm}\label{coeffbound}
The coefficients $c_h$ of the function $F_V$ in the previous theorem satisfy the estimate 
$$ c_h \leq C\,  e^{ (\sqrt{2c/3}+\varepsilon)\, \pi \, \sqrt{h}}$$ with
a positive constant $C$ and an arbitrary small $\varepsilon>0$. 
\end{thm}
\pf It was shown by Ng and Schauenburg~\cite{NS} that modular tensor categories define a projective representation $\overline{\rho}$
of ${\rm SL}(2,\Z)$ with a principal congruence subgroup of some level $N$ in its kernel. 
From this it can be deduced that the same holds for the representation $\rho$ on the space of genus one $1$-point correlation functions of a strongly-rational
vertex operator algebra~\cite{DLN}. Thus the characters $\chi_{M_\lambda}$ of irreducible $V$-modules $M_\lambda$ are weakly holomorphic modular functions for
some congruence subgroups of level~$N$ with Fourier expansions of the form 
$$\chi_V=\sum_{n\in\Q,\ n\geq -\frac{c}{24}} a(n)\,q^n$$
in all cusps.
It now follows from the Hardy-Ramanujan-Rademacher type series, that for the coefficients $a(n)$ of the expansions in all cusps one has an estimate
$$|a(n)|\leq C_1\, I_1(4\pi\sqrt{n c/24})\, e^{\epsilon \sqrt{n}}$$
where $C_1$ is a positive constant, $\varepsilon>0$ is arbitrarily small and
$I_1(x)$ is the modified Bessel function of the first kind; cf.~\cite{Bo-estimate}, Lemma~5.3.
The estimate $I_1(x)=\frac{e^x}{\sqrt{2\pi x}}(1+O(1/x))$ for large $x$ implies now the theorem. \qed

\smallskip

For general central charge~$c$, we have to consider the decomposition of the $V$-modules into irreducible Virasoro modules. 
The characters of the Virasoro algebra modules $M(c,h)$  and $\overline{M}(c,0)$ are given by
$$\chi_{c,h}=\chi_{M(c,h)}=\frac{q^{-c/24+h}}{\prod_{n=1}^\infty (1-q^n)} \hbox{\ and\ } 
\overline{\chi}_{c}=\chi_{\overline{M}(c,0)}=\frac{q^{-c/24}}{\prod_{n=2}^\infty (1-q^n)}=(1-q)\,\chi_{c,0},$$
respectively. Thus the following holds:
\begin{thm}\label{stronglypositive}
For a unitary strongly-rational vertex operator algebra $V$, the function 
$F_V=\frac{1}{2}\left(\chi_V+\chi_V|S\right)$ is an 
$S$-invariant function on the upper half-plane having the expansion
$$F_V=c_0\, \overline{\chi}_{c} +\sum_{h>0} c_h\,\chi_{c,h}$$
with coefficients $c_h\geq 0$ and $c_0>0$. 
\end{thm}

It is clear that if $V$ is a unitary strongly-rational vertex operator with minimal dual weight $\mu^*_V$ 
then the $q$-expansion of $F_V$ has the form $F_V=c_0\, \overline{\chi}_{c} +\sum_{h\geq \mu_V^*} c_h\,\chi_{c,h}.$


\section{Numerical semidefinite programming bounds}

We apply the conformal bootstrapping method to a function $G_V$ obtained from the character of $V$ 
to find  numerical upper bounds for $\mu_V^*$ for fixed $c$.

\bigskip

By using the Dedekind eta-function $\eta=q^{1/24}\,\prod_{n=1}^\infty (1-q^n)$ 
the characters of the Virasoro algebra modules $M(c,h)$  and of $\overline{M}(c,0)$ can be rewritten as
$$\chi_{c,h}=\frac{q^{1-\frac{c}{24}+h}}{\eta} \hbox{\ and\ } \overline{\chi}_{c}=\frac{q^{1-\frac{c}{24}}(1-q)}{\eta}=(1-q)\,\chi_{c,0},$$
respectively.
\medskip
Since the eta-function has the transformation property 
$\eta(-1/\tau)=(\tau/i)^{1/2}\eta(\tau)$, the following result 
follows directly from Theorem~\ref{stronglypositive}.
\begin{prop}\label{GV}
Let $V$ be a unitary strongly-rational vertex operator algebra of central charge~$c>1$ and let
\begin{eqnarray*}
G_V & =   & \frac{1}{2}\left(\frac{\tau}{i}\right)^{1/4} \eta(\tau)\left(\chi_V+\chi_V|S\right) \\
  &= &   c_0\,\left(\frac{\tau}{i}\right)^{1/4}(1-q)\,q^{(1-c)/24}+\sum_{h> 0} c_h\,\left(\frac{\tau}{i}\right)^{1/4}\,q^{(1-c)/24+h}.
\end{eqnarray*}
Then one has $G_V=G_V|S$ and $G_V$ has the given expansion with non-negative coefficients $c_h$ and $c_0>0$.
If~$V$ has minimal dual conformal weight $\mu^*_V$ then $c_h=0$ for all $h$ in the 
open interval from $0$ to $\mu^*_V$.
\end{prop}

We will show that these properties of $G_V$ are enough to obtain upper bounds for~$\mu^*_V$.
The approach is a slight variation to the one used for lattices. 

We restrict $\tau$ to the positive imaginary axis and write $\tau=i\,t$ with positive real $t$.
Then $\overline{G}_V(t)=G_V(it)$ is a real-valued and real analytic function and satisfies 
$$\overline{G}_V\left(\frac{1}{t}\right)=\overline{G}_V(t).$$

By writing $t=e^z$ we see that $\overline{G}_V(e^z)$ is an even function and so 
\hbox{$\partial_z^n(\overline{G}_V(e^z))\bigm\vert_{z=0}=0$} for odd $n$. From $\partial_z=t\, \partial_t$ it
follows that the functional ${\cal D}$ defined by
$${\cal D}[\overline{G}_V]=\sum_{m=0}^N a_{2m+1}\,(t\,\partial_t)^{2m+1}(G_V)\Bigm\vert_{t=1}$$
annihilates $\overline{G}_V$, i.e.\, ${\cal D}[\overline{G}_V]=0$.

For the terms in the expansion of $\overline{G}_V$ as in Proposition~\ref{GV} with $h>0$ one has
$$(t\,\partial_t)^m\left(t^{1/4}e^{2\pi(-(1-c)/24 - h)t}\right)\Bigm\vert_{t=1}=p_m(h)\,e^{2\pi(-(1-c)/24- h)}$$
with a real polynomial $p_m$ of degree~$m$
and so 
$${\cal D}\left[t^{1/4}e^{2\pi( -(1-c)/24 - h)t}\right]=p(h)\,e^{2\pi((c-1)/24-h)}$$
where $p=\sum_{m=0}^N a_{2m+1}\,p_{2m+1}$ is a polynomial of degree $2N+1$.
Similarly, we can evaluate the first term of the sum for $G_V$ and get for 
$${\cal D}\left[t^{1/4}\,e^{-2\pi (1-c)/24 \cdot t}\,(1-e^{-2\pi t})\right]$$
a sum $r=\sum_{m=0}^N a_{2m+1}\,r_{2m+1}$  where the real numbers $r_{2m+1}$ depend only on 
the central charge $c$.

Assume we can find real numbers $a_{2m+1}$, $m=0$, $\ldots$, $N$, such that $r >0$ and $p(h)\geq 0$ for $h\geq \Delta$.
Then one has
$$0 ={\cal D}[\overline{G}_V]=c_0\,r+\sum_{h>0 } c_h\,p(h)\,e^{2\pi(-(1-c)/24- h)}.$$
Since $c_0\, r$ is positive and $c_h\,p(h)\,e^{2\pi(-(1-c)/24- h)}$ is non-negative for $h\geq \Delta$, at least one of the $c_h$ for  $0<h< \Delta$
has to be nonzero. Thus we have found an upper estimate $\mu^*_V<\Delta$ which must hold for all vertex operator algebras $V$.

\smallskip

We used again the solver SDPB~\cite{sdpb} for this polynomial type semidefinite programming problem.
We have chosen $r$ as the objective function to be maximized and $p(x+\Delta)\geq 0$ for $x\geq 0$ as the polynomial constrained.
For fixed central charge $c$, the approximately smallest values of $\Delta$ for which a positive solution for the objective function exists seem again to converge 
for increasing degree $2N+1$ of the polynomial~$p$. The results of the computation for selected central charges are shown in 
Table~\ref{voabounds}.

\begin{table}\caption{Semidefinite programming bounds for $\mu_V^*$}\label{voabounds}

\def\ha{\frac{1}{2}}

$$\small\begin{array}{c|*{10}{l}}
c    & \frac{8}{7}  &  2  & 4       & 7       & 8          & 16      &  23\ha   & 24      & 32       & 48    \\[1mm] \hline

N=12 & 0.517   &  0.592   & 0.736   & 0.937   & 1.0022     & 1.521   &  2.005   & 2.037   &  2.555   & 3.603 \\
N=24 & 0.5165  &  0.5919  & 0.7353  & 0.9351  & 1.000089   & 1.5082  &  1.9743  & 2.0052  &  2.4978  & 3.4811\\
N=36 & 0.51646 &  0.59189 & 0.73523 & 0.93497 & 1.000008   & 1.50712 &  1.97044 & 2.00107 &  2.48828 & 3.45357 

\end{array}
$$ 
\end{table}

We summarize the calculations as:
\begin{thm}
For unitary strongly-rational vertex operator algebras of central charges 
$c=\frac{8}{7}$, $2$, $4$, $7$, $8$, $16$, $23\frac{1}{2}$, $24$, $32$ and $48$ the minimal dual conformal weight
has the upper bound given by the entry in row $N=36$ of Table~3. 
\end{thm}

\medskip

The computations suggest that for increasing $N$ the obtained bounds $\mu^*_V$  for central charges $c=8$ and $c=24$ converge
to $1$ and $2$, respectively. This would be the best possible upper bounds since 
the vertex operator algebra associated to the affine Kac-Moody algebra $\widetilde{E}_8$ at level~$1$ and the moonshine module $V^\natural$
realize these bounds. 

\smallskip

In the next section, we will show that the exact programming bounds 
are indeed $\mu_V^*\leq 1$ for $c=8$ and $\mu_V^* \leq 2$ for $c=24$.

\medskip

It seems that for all central charges $c$ not in the interval $[8,24]$ one always gets a bound for $\mu^*_V$ better than $\frac{c}{16}+\frac{1}{2}$.
Analogous to the lattice situation, one expects an asymptotic upper bound $\mu^*_V\leq c/\alpha$ with a constant
$\alpha$ slightly larger than $16$; cf.~\cite{HMR} where a slightly different programming bound for conformal field theories is
considered.

\smallskip

As discussed in the proof of Theorem~\ref{coeffbound}, the character $\chi_V$ is a modular function for $\Gamma(N)$ for a certain level~$N$.
Similarly as in the situation for lattices, one may hope that there is a certain distinguished subspace of Teichm\"uller  modular functions to which
for maximal $V$ the character $\chi_{V}(-1/\tau)$ must belong and  which has the dimension $c/24+O(1)$ for large $c$. 
The results mentioned in the last section show that this asymptotic bound holds for vertex operator algebras with $N=1$ and $N=2$.

\medskip

{\bf Problem:} Does there exist a linear asymptotic {\em lower bound,\/} i.e.\ are there unitary strongly-rational vertex operator algebras with $\mu_V^*\geq c/\beta$ for
a positive constant $\beta$ and arbitrarily large central charges $c$?


\def\FF{ \phantom{\!|}_2F_1 }
\def\ha{ {\textstyle\frac{1}{2}} }

\section{Exact bounds for central charges  $8$ and $24$}

We start by reviewing properties of the $\lambda$-function.

\smallskip

The modular $\lambda$-function can be defined as the ratio $\displaystyle\lambda=\frac{e_3-e_2}{e_1-e_2}$,
where the $e_i$ are the values of the Weierstra\ss{} $\wp$-function for the lattice
$L=\Z\,\omega_1+\Z\,\omega_2$ at the two-division points
of the elliptic curve $\C/L$:
$$e_1=\wp_L(\frac{\omega_1}{2}), \qquad e_2=\wp_L(\frac{\omega_2}{2}), \qquad e_3=\wp_L(\frac{\omega_1+\omega_2}{2}).$$
The group of basis changes of $L$ acts on the set of two-division points
as the full permutation group $S_3$.

As a function of $\tau=\omega_2/\omega_1$, the $\lambda$-function defines an
$S_3$-equivariant biholomorphic map from the modular curve
$\overline{\H/\Gamma(2)}$ to the Riemann sphere $\widehat\C$ mapping the three 
cusps $i\infty$, $0$ and $1$ of $\overline{\H/\Gamma(2)}$ to the points $0$, $1$ and $\infty$, respectively.

Here, the action of $S_3$, is given by the action of $\Gamma/\Gamma(2)={\rm PSL}(2,\F_2)$ 
on $\overline{\H/\Gamma(2)}$ and on $\widehat\C$ it
is given by the group of biholomorphic transformations of $\widehat\C\setminus \{0,1,\infty \}$.
Explicitly, the action of the generators $S$ and $T$ of $\Gamma$ are 
are described in the following table:
$$\begin{array}{cccc}
{\rm representative} & {\rm order } & {\rm action\ on}\ \H & {\rm action\ on}\ \widehat\C\setminus \{0,1,\infty \} \\ \hline \\[-2mm]
S        & 2 &  \tau \mapsto -1/\tau     & z\mapsto 1-z \\[2mm]
T        & 2 &  \tau \mapsto\tau+1        &   z\mapsto \displaystyle \frac{z}{z-1}  \\[3mm]
\end{array}
$$
The subgroup $\Gamma(2)\subset \Gamma={\rm PSL}(2,\Z)$ is isomorphic to the free group with two generators and hence the upper
half-plane forms an universal cover of $\H/\Gamma(2)$.

We use the notation $f|A$ for the induced action of an element $A$ of the complex group algebra  ${\bf C}[\Gamma]$ on an element $f$ 
of the space of holomorphic functions on the appropriate domain.

The expansion of $\lambda$ in $q=e^{2\pi i \tau}$ is 
$$\lambda(\tau)= 16\, q^{1/2} -128\, q + 704\,q^{3/2} -3072\,q^2 + 11488\, q^{5/2} - 38400\,q^3+ \cdots .$$

\smallskip

An inverse of the modular $\lambda$-function on the universal cover of $\widehat\C\setminus \{0,1,\infty \}$ is given by 
$$\lambda^{-1}(z)=i\,\frac{ \FF(\ha,\ha;1;1-z)}{ \FF(\ha,\ha;1;z) }.$$
Here, the hypergeometric series
$$\FF(\ha,\ha;1;z)=\sum_{n=0}^\infty \frac{\bigl(\ha\cdot\frac{3}{2}\cdots (n-\ha)\bigr)^2}{1\cdot 2 \cdots n}\,\frac{z^n}{n!}$$
converges for $|z|<1$ and defines by analytic continuation a function on the universal cover of $\widehat\C\setminus \{0,1,\infty \}$.
We usually consider the branch which is defined
on the doubly sliced complex plane  $U=\C\setminus \left( (-\infty,0]\cup [1,\infty) \right)$.
The function  $\FF(\ha,\ha;1;z)$ can also be expressed in terms of the elliptic integral
$$ \int_0^1 \frac{dt}{\sqrt{t(1-t)(1-tz)}}=\pi\, \FF(\ha,\ha;1;z). $$


\medskip


\smallskip

We use $\lambda^{-1}$ to define for real $c$ and $h$ on $U$ the function
$$\displaystyle H_{c,h}(z)=\left(\frac{1}{z(1-z)}\right)^{c/24} e^{2\pi i\, (-\frac{c}{24}+h)\, \lambda^{-1}(z)}.$$
We denote by
$H^+_{c,h}(z)$ and $H^-_{c,h}(z)$ the analytic continuations of $H_{c,h}(z)$ from $U$  to 
the branch cut $(-\infty,0)$ coming from above and below, respectively.

\begin{lem}\label{Htransform}
One has $H^\pm_{c,h}(z\vert T)=e^{\pm 2 \pi i h}\,(z-1)^{c/8}\,H_{c,h}(z)$ for $z$ in $(0,1)$.
\end{lem}

\pf First, we note 
$$z^{c/8}\left(\frac{1}{z(1-z)}\right)^{c/24}=\left(\frac{z^2}{1-z}\right)^{c/24}.$$
We will check the monodromy property of 
$$\left(\frac{z^2}{1-z}\right)^{c/24}\, q^{-\frac{c}{24}+h}$$
around $0$.
The power series expansion of $\lambda^{-1}(z)$ gives for $q=e^{2 \pi i \tau}$ as a function of $z$ the series
$$q(\lambda^{-1}(z))=\frac{z^2}{256}+\frac{z^3}{256}+\frac{29\,z^4}{8192}+\cdots. $$
Thus we have the expansion
\begin{eqnarray*}
   \left(\frac{z^2}{1-z}\right)^{c/24}\, q^{-\frac{c}{24}+h} & = & 
     \left(\frac{z^2}{1-z}\right)^{c/24}  \left( \frac{z^2}{256}+\frac{z^3}{256}+\frac{29\,z^4}{8192}+\cdots\right)^{-c/24+h} \\
 & = &  2^{\frac{c}{3}-8h}\,z^{2h}\left(\frac{1}{1-z}\right)^{c/24}\left(1+z+\frac{29}{32}\,z^2+\cdots\right)^{-c/24+h} 
\end{eqnarray*} 
Since one has $z|T=-z +O(z^2)$,
it follows that $z^{c/8}\,H^\pm_{c,h}(z)\vert T=e^{\pm 2 \pi i h}\,z^{c/8}\,H_{c,h}(z)$. 
From this one gets 
$$H^\pm_{c,h}(z)\vert T=e^{\pm 2 \pi i h}\,\frac{z^{c/8}}{(z^{c/8})\vert T}\,H_{c,h}(z)=e^{\pm 2 \pi i h}\,(z-1)^{c/8}\,H_{c,h}(z).\qquad\qquad \qed$$

\begin{lem}\label{Hasymptotics}
The function $H_{c,h}(z)$ has the following asymptotic behavior:
\begin{itemize}
\item[] For $z\longrightarrow 0$ one has $H_{c,h}(z)=O(z^{-c/8+2h})$.
\item[] For $z\longrightarrow 1$ one has $H_{c,h}(z)=O((1-z)^{-c/24})$.
\item[] For $z\longrightarrow \infty$ one has $H_{c,h}(z)=O(z^{-c/12})$.
\end{itemize}
\end{lem}
\pf  Since for $z\longrightarrow 0$ one has $q(\lambda^{-1}(z))=z^2/256+O(z^3)$ one gets 
$$H_{c,h}(z)=O(z^{-c/24})\cdot O(z^{-c/12+2h})=O(z^{-c/8+2h}).$$
For $z\longrightarrow 1$ one gets $q\longrightarrow 1$ and so
$$H_{c,h}(z)=O((1-z)^{-c/24})\cdot O(1)=O((1-z)^{-c/24}).$$
For $z\longrightarrow \infty$ one gets  $q\longrightarrow -1$
and so
$$H_{c,h}(z)=O(z^{-c/12})\cdot O(1)=O(z^{-c/12}).\qquad  \qed$$

\medskip

We need the two ``magic'' rational functions
$$ A_8(z)=\frac{(1-z)(2z^2+z+2)}{z^2}$$
$$ A_{24}(z)=\frac{(1-z)^3(2\,z^2+3\,z+2)}{z^2}.$$
For the rest of this section let $c=8$ or $c=24$. 
One easily checks:
\begin{lem}\label{Qprop}
The two functions $A_c(z)$ are positive on the open interval~$(0,1)$ and
satisfy the functional equation
$$A_c(z)+A_c(z\vert S)-(1-z)^{\frac{c}{8}-2}A_c(z\vert T)=0  .$$
They have the asymptotic behavior
$A_c(z)=O( (1-z)^{c/8})$ for $z\longrightarrow  1$ and $A_c(z)=O(z^{-2})$ for  $z\longrightarrow  0$.
\end{lem}

We set $B_c(z)=-(1-z)^{\frac{c}{8}-2}\,A_c(z\vert T)$.
Explicitly, one has
$$B_8(z)=-\frac{5(z-1)z+2}{(z-1)^2z^2}$$ 
$$B_{24}(z)=-\frac{7(z-1)z+2}{(z-1)^2z^2}.$$
We note:
\begin{lem}\label{Rprop}
The two functions $B_c(z)$ are $S$-invariant and have the asymptotic behavior $B_c(z)=O(z^{-2})$ for  $z\longrightarrow  \infty$.
\end{lem}

\medskip

We use the rational functions $A_c$ and $B_c$ to define the contour integral
$${\cal D}_c[\varphi]=\int_{ \frac{1}{2}}^1 \varphi(z)\, A_c(z)\, dz +
         \frac{1}{2}\,  \int_{\frac{1}{2}}^{\frac{1}{2}+i\infty} \varphi(z)\, B_c(z)\, dz$$
which provides a linear functional on the  subspace of  holomorphic functions $\varphi$
on $U$ for which the integral exists.

\begin{prop}\label{existence}
The integral ${\cal D}_c[\varphi]$ exists for $\varphi=H_{c,h}(z)\big\vert({\rm id}-S)$ for all non-negative real numbers $h$.
\end{prop}
\pf
This follows from the estimates given in Lemma~\ref{Hasymptotics}, Lemma~\ref{Qprop} and Lemma~\ref{Rprop}. 
The integral $\int_{1/2}^1 H_{c,h}(z) \, A_c(z)\, dz$ exists since  for $z\longrightarrow 1$ one has
$$H_{c,h}(z)\,A_c(z)
=O((1-z)^{-\frac{c}{24}})\cdot O( (1-z)^{\frac{c}{8}}) = O((1-z)^{ \frac{c}{12}}).  $$
The integral $\int_{1/2}^1 H_{c,h}(z|S) \, A_c(z)\, dz$ exists since for $z\longrightarrow 1$ one has
$$H_{c,h}(1-z)\,A_c(z)
=O((1-z)^{-\frac{c}{8}+2h})\cdot  O( (1-z)^{\frac{c}{8}}) = O((1-z)^{2h}).  $$
The integrals $\int_{1/2}^{1/2+i\infty} H_{c,h}(z) \, B_c(z)\, dz$ and  $\int_{1/2}^{1/2+i\infty} H_{c,h}(z|S) \, B_c(z)\, dz$  exist since for $z\longrightarrow \infty$ one has
for $H_{c,h}(z)\,B_c(z)$ and $H_{c,h}(z|S)\,B_c(z)$ the asymptotic behavior
$$\qquad\qquad\qquad\qquad\qquad\qquad  O(z^{-\frac{c}{12}})\cdot O( z^{-2}) = O(z^{-\frac{c}{12}-2}).\qquad\qquad\qquad \qquad\qquad\qquad \qed  $$

\medskip

We study now for $h\geq 0$ the function
$$f_c(h)={\cal D}_c\left[ H_{c,h}(z)\big\vert({\rm id}-S)\right].$$
\begin{thm}\label{contour}
The function $f_c(h)$ is real-valued and non-negative for $\displaystyle h\geq \frac{c}{16}+\frac{1}{2}$ and 
vanishes for  $\displaystyle h= \frac{c}{16}+\frac{1}{2}+n$, $n=0$, $1$, $2$, $\ldots$.
\end{thm}
\pf  We claim that for  $h> \frac{c}{16}+\frac{1}{2}$ a contour deformation argument will give
\begin{equation}\label{sineintegral}
f_c(h)=2\, \sin^2(h\,\pi)\, \int_0^1 H_{c,h}(z)\, A_c(z)\, dz.
\end{equation}
\smallskip

To prove the claim, we consider
$$F_{c,h}(z)=H_{c,h}(z)+\frac{1}{2}\,(1-z)^{-c/8}\,\left(H^+_{c,h}(z|T)+H^-_{c,h}(z|T) \right)$$
By Lemma~\ref{Htransform} we have $H^\pm_{c,h}(z|T)=e^{\pm 2\pi i h}(z-1)^{c/8}\,H_{c,h}(z)$.
It follows that $F_{c,h}(z)=2\,\sin^2(h\,\pi)\, H_{c,h}(z)$ for $z$ in $(0,1)$.
Integration gives
\begin{equation}\label{sineformula}
\int_0^1 F_{c,h}(z)\, A_c(z)\,dz=2\,\sin^2(h\,\pi )\int_0^1 H_{c,h}(z) \, A_c(z)\, dz
\end{equation}
provided the integral exists. 
This is the case for $h>c/16+1/2$ since from Lemma~\ref{Hasymptotics} it follows that 
$H_{c,h}(z)\, A_c(z)=O(z^{-c/8+2h-2})=O(z^{-1+s})$ for $z\longrightarrow 0$ and an $s>0$.

We now evaluate $\int_0^1 F_{c,h}(z)\, A_c(z)\,dz$ by contour deformation and 
by using Lemma~\ref{Qprop} and  Lemma~\ref{Rprop}.
\begin{eqnarray}\label{deformation} \nonumber
\int_0^1 F_{c,h}(z)\, A_c(z)\,dz \!\!\!\!\!\!\!\!\!\!\!\!\!\!\!\!\!\!\!\!\!\!\!\!\!\!\!\!\!\!\!\!\!\!\!\!\!\!\!\! \\  \nonumber
& = &\int_0^1 H_{c,h}(z)\, A_c(z)\,dz +\frac{1}{2}\int_0^1\,(1-z)^{-c/8}\,H^+_{c,h}(z|T) A_c(z)\,\,dz\\  \nonumber
&& \qquad \qquad \qquad \qquad \qquad \qquad \qquad \qquad + \frac{1}{2}\int_0^1\,(1-z)^{-c/8} H^-_{c,h}(z|T) A_c(z)\, \,dz  \\ \nonumber
& = &  \int_{1/2}^1 H_{c,h}(z)\, A_c(z)\,dz-\int_{1/2}^0 H_{c,h}(z)\, A_c(z)\,dz\\ \nonumber
& & \quad\qquad +\frac{1}{2}\int_0^{1/2+i\infty}\,H_{c,h}(z)\,B_c(z)\,dz+\frac{1}{2}\int_{0}^{1/2-i\infty}\,H_{c,h}(z)\, B_c(z)\,dz\\ \nonumber
& = &  \int_{1/2}^1 H_{c,h}(z)\, A_c(z)\,dz-\int_{1/2}^0 H_{c,h}(z)\, A_c(z)\,dz\\ \nonumber
& & \quad\qquad +\frac{1}{2}\int_{1/2}^{1/2+i\infty}\,H_{c,h}(z)\,B_c(z)\,dz+\frac{1}{2}\int_{1/2}^{1/2-i\infty}\,H_{c,h}(z)\, B_c(z)\,dz\\ \nonumber
& & \quad \quad \qquad \quad \qquad \quad \qquad \quad \qquad \quad \qquad \quad \qquad \quad \qquad-\int_{1/2}^0\,H_{c,h}(z)\,B_c(z)\,dz\\ \nonumber
& = & \int_{1/2}^1 H_{c,h}(z)\, A_c(z)\,dz+\int_{1/2}^0 H_{c,h}(z)\, A_c(z\vert S)\,dz+\\ \nonumber
&& \qquad \qquad \frac{1}{2}\,\int_{1/2}^{1/2+i\infty} H_{c,h}(z)\, B_c(z) \,dz+ \frac{1}{2}\,\int_{1/2}^{1/2-i\infty} H_{c,h}(z)\, B_c(z) \,dz \\ \nonumber
& = & \int_{1/2}^1 H_{c,h}(z)\big\vert ({\rm id} -S)\, A_c(z)\,dz + \frac{1}{2}\,\int_{1/2}^{1/2+i\infty} H_{c,h}(z) \big\vert ({\rm id} -S)\, B_c(z)\,dz\\ 
& = & f_c(h).
\end{eqnarray}
Taken together, equations~(\ref{sineformula}) and~(\ref{deformation}) prove the claim.
\medskip
The three factors 
\vspace{-3mm}
$$ e^{2\pi i\, (-\frac{c}{24}+h)\, \lambda^{-1}(z)},\qquad  \frac{1}{(z(1-z))}^{c/24}  \qquad \hbox{and}\qquad A_c(z)$$
are all real valued  and non-negative on the interval $(0,1)$, cf.\ Lemma~\ref{Qprop}. It follows then from
equation~(\ref{sineintegral}) that  $f_c(h)$ is real-valued and non-negative for $h> \frac{c}{16}+\frac{1}{2}$ and 
vanishes for  $h=\frac{c}{16}+\frac{1}{2}+n$, $n=1$, $2$, $\ldots$.

\smallskip

For   $h=\frac{c}{16}+\frac{1}{2}$ we note that the integral $\int_0^1H_{c,h}(z)\, A_c(z)\,dz$ develops a simple pole if $h$ approaches $\frac{c}{16}+\frac{1}{2}$,
but the factor $\sin^2(h\,\pi)$ provides a double zero, so that the product of both as an analytic function in $h$ must be zero.
\qed

\begin{prop}\label{absoluteexistence}
For $h\geq h_0>\frac{c}{16}+\frac{1}{2}$, the integral ${\cal D}_c\left[ H_{c,h}(z)\big\vert({\rm id}-S) \right]$ exists as an absolute integral which is bounded by 
$$ C\, e^{ (-2\,\sqrt{(c+12)/6}+\varepsilon) \pi  \sqrt{h}}$$ 
with a constant $C>0$ and an arbitrarily small constant $\varepsilon>0$.
\end{prop}

\pf
For $h>\frac{c}{16}+\frac{1}{2}$, we can use the integral representation as in equation~(\ref{sineformula}). 
We split the contour into two parts for which we provide individual estimates.

\smallskip

We have 
\begin{eqnarray*}
\int_{1/2}^1 H_{c,h}(z) \, A_c(z)\, dz & = & \int_i^0 e^{2\pi i\,( -\frac{c}{24}+h)\,\tau} \,\frac{A_c(\lambda)\, \lambda'(\tau)}{(\lambda(1-\lambda))^{c/24}}\,\,d\tau,\\
& = & \int_i^{i\infty} e^{2\pi i\,( -\frac{c}{24}+h)\,(-\frac{1}{\tau})} \,\frac{A_c(1-\lambda)\, (-\lambda'(\tau))}{(\lambda(1-\lambda))^{c/24}}\,d\tau.
\end{eqnarray*}
By using the $q$-expansion of $\lambda$ one finds 
$$\frac{A_c(1-\lambda)\, (-\lambda'(\tau))}{ (\lambda(1-\lambda))^{c/24}}=C_0\,q^{\frac{1}{2}+\frac{c}{24}}(1+O(q)).$$
For $\tau\longrightarrow i\infty$, a series expansion of a weakly holomorphic modular function of any weight is dominated by the first term. 
Thus we get the estimate
$$ \int_{1/2}^1 |H_{c,h}(z) \, A_c(z)|\, dz \leq C_1 \int_1^{\infty}e^{-2\pi (-\frac{c}{24}+h)/t}\, e^{- 2\pi \,(\frac{1}{2}+\frac{c}{24}) t}\leq   C_1 \int_0^{\infty}e^{-2\pi (-\frac{c}{24}+h)/t}\, e^{- 2\pi \,(\frac{1}{2}+\frac{c}{24}) t}$$
$$\qquad\qquad\qquad\qquad\qquad  =2\,\sqrt{\frac{-c+24\,h}{c+12}}\, K_1(\sqrt{(c+12)(-c+24\,h)}\,\pi/6)\leq C_2\, e^{-(2\,\sqrt{(c+12)/6}-\varepsilon)\,\pi\,\sqrt{h}} $$
where $K_1(x)$ is the modified Bessel function of the second kind which has the asymptotic
$$K_1(x)=\sqrt{\frac{\pi}{2x}}\, e^{-x}\, \left(1+O\Bigl(\frac{1}{x}\Bigr)\right)\qquad \hbox{for\ } x\longrightarrow \infty.$$

On the other hand,
\begin{eqnarray*}
\int_{0}^{1/2} H_{c,h}(z) \, A_c(z)\, dz & =  & - \int_i^{i\infty}  e^{2\pi i\,( -\frac{c}{24}+h)\,\tau} \,\frac{A_c(\lambda)\,\lambda'(\tau)}{(\lambda(1-\lambda))^{c/24}}\,\,d\tau.
\end{eqnarray*}
The $q$-expansion
$$\frac{A_c(\lambda)\, \lambda'(\tau)}{ (\lambda(1-\lambda))^{c/24}}=C_2\,q^{-\frac{1}{2}-\frac{c}{48}}(1+O(q))$$
gives the estimate
$$ \int_{0}^{1/2} |H_{c,h}(z) \, A_c(z)|\, dz \leq C_3 \int_1^{\infty}e^{-2\pi (-\frac{c}{24}+h)\,t}\, e^{ 2\pi(\frac{1}{2}+\frac{c}{48})\, t}=\frac{C_3\,e^{2(\frac{c}{16}+\frac{1}{2}-h)\pi}}{2\,(h-\frac{c}{16}-\frac{1}{2})\pi}. $$
Taken the two estimates together, the proposition follows. \qed

\medskip


We consider sums of the form 
\begin{equation}\label{Z}
 Z(\tau)=c_0\,q^{-c/24} + \sum_{h>0} c_h\, q^{-c/24+h}
\end{equation}
where the number of non-zero real $c_h$ for $h$ in an interval $[0,u]$ is bounded by a linear function in $u$.
We also assume that the coefficients $c_h$ satisfy the estimate 
$$0\leq c_h \leq C'\,  e^{ (\sqrt{2c/3}+\varepsilon') \pi  \sqrt{h}}$$ with
a positive constant $C'$ and an arbitrarily small $\varepsilon'>0$. 
Then the sum converges locally uniformally to a holomorphic function on the
upper half-plane.



\begin{prop}\label{mainestimate}
If $Z(\tau)$ is $S$-invariant then there exists either at least one non-zero coefficient $c_h$  with $0<h<\frac{c}{16}+\frac{1}{2}$ 
or there is no such non-zero coefficient $c_h$ but
all coefficients $c_h$ with non-integral $h$ are vanishing.
\end{prop}

\pf Assume that all  $c_h$ with $0<h<\frac{c}{16}+\frac{1}{2}$ vanish.
From equation~(\ref{Z}) we obtain
\begin{equation}\label{rightside}
 \left(\frac{1}{z(1-z)}\right)^{c/24}Z(\lambda^{-1}(z))\big\vert({\rm id}-S) = c_0\,H_{c,0}(z) \big \vert({\rm id}-S)   +
 \sum_{h\geq c/16 +1/2} c_h\, H_{c,h}(z) \big \vert({\rm id}-S) . 
\end{equation}

From the $S$-invariance of $Z(\tau)$ it follows that 
$$ \left(\frac{1}{z(1-z)}\right)^{c/24}Z(\lambda^{-1}(z))\big\vert({\rm id}-S)= 0.$$
\smallskip
We like to apply ${\cal D}_c$ to equation~(\ref{rightside}) and evaluate the right hand sum
by applying it to the individual terms.

By the Fubini-Tonelli theorem (using the counting measure on ${\bf N}$) we are allowed to do this if we can show
that the individual integrals exist absolutely and the sum of the absolute integrals is finite.

By Theorem~\ref{existence}, the functional is well-defined for each individual term of the sum.
By Theorem~\ref{absoluteexistence},  the absolute integrals exist for $h>\frac{c}{16}+\frac{1}{2}$.
For the sum of the absolute integrals,  Theorem~\ref{absoluteexistence}
together with the assumption on the coefficients $c_h$ gives
$$ C'\,  e^{ (\sqrt{2c/3}+\varepsilon) \pi \sqrt{h}}\cdot C\, e^{ (-2\,\sqrt{(c+12)/6}+\varepsilon') \pi  \sqrt{h}} =  C''\,e^{(\sqrt{2/3}\,(\sqrt{c}-\sqrt{c+12})+\varepsilon'')\,\pi\,\sqrt{h} }$$
and so the sum converges absolutely.

\smallskip
 
Applying $ {\cal D}_c$ to equation~(\ref{rightside}) results now  in a contradiction to Theorem~\ref{contour} 
unless all $c_h$ with non-integral~$h$ are vanishing or
the first term of the right hand side of  equation~(\ref{rightside}) is negative. However, this is impossible:
Let $j$ denote the modular $j$-function.
By taking $Z(\tau)=j(\tau)^{1/3}$ if $c=8$ and  $Z(\tau)=j(\tau)-744$ if $c=24$ we see that the first term must be zero since only
$c_h$ with integral $h$ (and $h\not=1$ for $c=24$) occur in the sum decomposition of $Z(\tau)$ for these two cases and 
$Z(\tau)$ is $S$-invariant. \phantom{cxxxxxxxxxxxxxxxxxxxxxxxx} \hfill \qed

\medskip

\begin{thm}
For the minimal dual weight $\mu_V^*$ of a unitary strongly-rational vertex operator algebra $V$ of central charge $c$,
one has the upper bounds $\mu_V^*\leq 1$ for $c=8$ and   $\mu_V^*\leq 2$ for $c=24$. 

If the bound is reached by a vertex operator algebra $V$ then $\chi_V=j^{1/3}$ for  $c=8$ and $\chi_V=j-744$ for $c=24$,
where $j$ denotes the modular $j$-function. In particular, $V$ is self-dual.
\end{thm}

\pf Let $F_V=\chi_V\vert ({\rm id} + S)$ where $\chi_V$ is the character of $V$. Then $F_V$ has by
Theorem~\ref{positive} a $q$-expansion with non-negative coefficients $c_h$ which by Theorem~\ref{coeffbound} are bounded by
$C'\,  e^{ (\sqrt{2c/3}+\varepsilon') \pi  \sqrt{h}}$ with constant $C'>0$ and $\varepsilon'>0$ arbitrarily small.
Since  $F_V=\chi_V\vert ({\rm id} + S)$ is also $S$-invariant, it satisfies all our assumptions for the function $Z(\tau)$ in
Proposition~\ref{mainestimate} and hence there is either a non-zero $c_h$ with $0<h<\frac{c}{16}+\frac{1}{2}$ or
all coefficients $c_h$ with non-integral $h$ vanish. 

Since  $\overline{\chi}_{c}(\tau)=q^{-c/24}(1+q^2+O(q^3))$, the vacuum vector ${\bf 1}$ provides no contribution to the coefficients $c_h$ in the sum
expansion of $F_V$ for $0<h<\frac{c}{16}+\frac{1}{2}$ and it follows that in the first case one has $\mu_V^*< \frac{c}{16}+\frac{1}{2}$.

\smallskip

For the second option of Proposition~\ref{mainestimate}, all coefficients $c_h$ with non-integral $h$ are vanishing.
This implies that $F_V(\tau)$ is also  $T$-invariant up to multiplication by a third root of unity (for $c=8$) or $T$-invariant (for $c=24$). 
The only functions $F_V(\tau)$ with these properties are multiples of $j^{1/3}$ and  $j-744$, respectively; cf.~\cite{Ho-dr}. 
Since $j^{1/3}=q^{-1/3}(1+248\,q+O(q^2))$ and  $j-744=q^{-1}+196884 \,q^1 +O(q^2)$, we conclude that in the second case  $\mu_V^*= \frac{c}{16}+\frac{1}{2}$.
This proves the first claim of the theorem.

\smallskip

For the second claim, we observe that if all coefficients $c_h$ with non-integral $h$ are vanishing then
$V$ would only have irreducible $V$-modules $M_\lambda$ with integral conformal weights. 
In this case, the action of $T$ on the space spanned by the characters of all irreducible $V$-modules is the multiplication with $e^{2\pi i c/8}$. Since $(ST)^3={\rm id}$ in $\Gamma$, it follows
that  $S^3=S$ acts trivially, i.e.\ the whole modular group $\Gamma$ acts by multiplications with powers of  $e^{2\pi i c/8}$ on all characters. It follows that already
$\chi_V=j^{1/3}$ (for $c=8$) or $\chi_V=j-744$ (for $c=24$) and $V$ must be self-dual.
\qed

\bigskip

For $c=8$, the only strongly-rational vertex operator algebra with the character $j^{1/3}$ is the affine $E_8$ level~$1$ Kac-Moody
vertex operator algebra.
For $c=24$, the only known vertex operator algebra with the character $j-744$ is the moonshine module and it
is conjectured that this is the only one~\cite{FLM}.


\begin{thebibliography}{9991}


\bibitem[BN]{BoeNe} S.\ B\"ocherer and G.\ Nebe, {\it On theta series attached to maximal lattices and their adjoints\/}, preprint, 
www.math.rwth-aachen.de/\~{}Gabriele.Nebe/papers/MAX.pdf (the version arXiv:0909.1184 is slightly different).

\bibitem[Bo1]{Bo-ur} R.\ E. Borcherds, {\it Vertex algebras, Kac-Moody algebras, and the Monster,} Proc. Nat. Acad. Sci. U.S.A. {\bf 83} (1986), 3068--3071. 

\bibitem[Bo2]{Bo-estimate} R.\ E. Borcherds, {\it Automorphic forms on $O_{s+2,2}(\R)$ and infinite products,} Invent.\ Math.\ {\bf 120} (1995), 161--213.

\bibitem[CKSMRV]{CKSMRV} H.\ Cohn, A.\ Kumar, S.\ D.\ Miller, D.\ Radchenko, M.\ Viazovska, {\it The sphere packing problem in dimension~$24$,}  Annals of Mathematics {\bf 85} (2017),  1017--1033, Arxiv:1603.06518.

\bibitem[CLY]{CLY} S.\ Collier, Y.-H.\ Lin, X.\ Yin, {\it Modular Bootstrap Revisited,\/} J. High Energ. Phys. (2018) 2018:61, 1--31,  ArXiv:1608.06241.

\bibitem[CS]{CoSl} J.~H. Conway and N.~J.~A. Sloane, \emph{{Sphere Packings, Lattices and
  Groups}}, second ed., Grundlehren der Mathematischen Wissenschaften Band 290,
  Springer-Verlag, New York, 1993.

\bibitem[DJ]{DJ} C.\ Dong and C. \ Jiang, {\it A Characterization of Vertex Operator Algebra $L(\frac{1}{2},0)\otimes L(\frac{1}{2},0)$,} Comm.\ Math.\ Phys.\ {\bf 296} (2010), 69--88.

\bibitem[DLN]{DLN} C.\ Dong, X.\ Lin and S.-H.\ Ng, {\it Congruence Property in Conformal Field Theory,} Alg.\ Numb.\ Theory {\bf 9} (2015), 2121--2166.

\bibitem[FLM]{FLM}
I.\  Frenkel, J.\ Lepowsky, and A. Meurman, {\it Vertex operator algebras and the Monster,} Pure and Applied Mathematics {\bf 134}, Academic Press, Inc., Boston, MA, 1988.  

\bibitem[FZ]{FZ}
 I.\ Frenkel and Y.\ Zhu, {\it Vertex operator algebras associated to representations of affine and Virasoro algebras,} Duke Math. J. {\bf 66} (1992), 123--168. 

\bibitem[FK]{FK} D.\ Friedan and C.\ Keller, {\it Constraints on 2d CFT partition functions,}  J. High Energ. Phys. (2013) 2013:180, 1--38, ArXiv:1307.6562.

\bibitem[FKS]{FKS} D.\ Friedan, A.\ Konechny, C.\ Schmid-Colinet, {\it Precise lower bound on Monster brane boundary entropy,} J. High Energ. Phys. (2013) 2013:99, 1--19,  ArXiv:1305.2122.


\bibitem[He]{He} S.\ Hellerman, {\it A universal inequality for CFT and quantum gravity,} S. J. High Energ. Phys. (2011) 2011: 130, 1--40, ArXiv:0902.2790.

\bibitem[HMR]{HMR} T.\ Hartman, D.\ Maz\'a\=c, L.\ Rastelli, {\it Sphere Packing and Quantum Gravity,}  ArXiv:1905.01319.

\bibitem[Ho1]{Ho-dr} G.\ H{\"o}hn, \emph{{Selbstduale Vertexoperatorsuperalgebren und das
  Babymonster}}, Ph.D. thesis, {U}niversit{\"a}t {B}onn, 1995, see: {B}onner
  {M}athematische {S}chriften {\bf 286}, arXiv:0706.0236.


\bibitem[Ho2]{Ho-extremal} G.\ H\"ohn, {\it Self-Dual Vertex Operator Superalgebras of Large Minimal Weight,} arXiv:0801.1822.

\bibitem[Ho3]{Ho-designs} G.\ H\"ohn, {\it Conformal Designs based on vertex operator algebras,} Advances in Mathematics {\bf 217} (2008), 2301--2335.

\bibitem[Hu]{Huang-tensor} Y.-Z. Huang, {\it Rigidity and modularity of vertex tensor categories,} Commun. Contemp. Math. {\bf 10} (2008) 871--911.

\bibitem[MS1]{McSl} F.\ J.\ MacWilliams, F. J. and  N.\ J.\ A.\ Sloane, 
{\it The theory of error-correcting codes,}
North-Holland Mathematical Library, Vol. 16. North-Holland Publishing Co., Amsterdam-New York-Oxford, 1977.

\bibitem[MS2]{MaSl} C.~L. Mallows and N.~J.~A. Sloane, \emph{{An upper bound for self-dual codes}},
Information and Control \textbf{22} (1973), 188--200.

\bibitem[NS]{NS}S.-H. Ng and P. Schauenburg,  {\it Congruence subgroups and generalized Frobenius-Schur indicators,}
Comm. Math. Phys. {\bf 300} (2010), 1--46.


\bibitem[R]{rains} E.\ Rains, {\it Bounds for Self-Dual Codes over ${\bf Z}_4$,} Finite Fields and their applications {\bf 6} (2000), 146--163.

\bibitem[SD]{sdpb} David Simmons-Duffin, {\it A Semidefinite Program Solver for the Conformal Bootstrap,} ArXiv:1502.02033.

\bibitem[V]{Vi} M.\ S.\ Viazovska, {\it The sphere packing problem in dimension~$8$,} Annals of Mathematics {\bf 85} (2017), 991--1015, Arxiv:1603.04246.

\bibitem[W]{3d-QG} E.\ Witten, {\it Three-Dimensional Gravity Revisited,} arXiv:0706.3359.

\bibitem[Zh]{Zhu-modular} Y. Zhu, {\it Modular invariance of characters of vertex operator algebras,} J. Amer. Math. Soc. {\bf 9} (1996) 237--302.

\end{thebibliography}
\end{document}